\documentclass[amssymb,prc,twocolumn,superscriptaddress,nofootinbib]{revtex4-2}
\usepackage{graphicx}
\usepackage{dcolumn}
\usepackage{adjustbox}
\usepackage{textcomp}
\usepackage{booktabs}
\usepackage{amsmath}
\usepackage[T1]{fontenc}
\usepackage[utf8]{inputenc}
\usepackage{color}
\usepackage{braket}
\usepackage{natbib}
\usepackage{mathrsfs}
\usepackage{dcolumn}

\usepackage{soul}
\usepackage[graphicx]{realboxes}
\usepackage{rotating}
\usepackage{tabularx} 
\begin{document}
	
\title[\empty]{Quadrupole dynamics of carbon isotopes and \boldmath$^{10}$Be}

\author{H.~Li}
\affiliation{Institute of Modern Physics, Chinese Academy of Sciences, Lanzhou 730000, China}
\affiliation{School of Nuclear Science and Technology, Lanzhou University, Lanzhou 730000, China}
\affiliation{School of Nuclear Physics, University of Chinese Academy of Sciences, Beijing 100049, China}
\affiliation{CAS Key Laboratory of High Precision Nuclear Spectroscopy, Institute of Modern Physics, Chinese Academy of Sciences, Lanzhou 730000, China}

\author{D.~Fang}
\affiliation{Institute of Modern Physics, Chinese Academy of Sciences, Lanzhou 730000, China}
\affiliation{School of Nuclear Physics, University of Chinese Academy of Sciences, Beijing 100049, China}

\author{H.~J.~Ong}
\affiliation{Institute of Modern Physics, Chinese Academy of Sciences, Lanzhou 730000, China}
\affiliation{School of Nuclear Physics, University of Chinese Academy of Sciences, Beijing 100049, China}
\affiliation{Joint Department for Nuclear Physics, Lanzhou University and Institute of Modern Physics, Chinese Academy of Sciences, Lanzhou 730000, China}
\affiliation{Research Center for Nuclear Physics, Osaka University, Ibaraki, Osaka 5670047, Japan}
\affiliation{RIKEN Nishina Center, Wako, Saitama 3510198,Japan}

\author{A.~M.~Shirokov}
\affiliation{Skobeltsyn Institute of Nuclear Physics, Lomonosov Moscow State University, Moscow 119991, Russia}

\author{J.~P.~Vary}
\affiliation{Department of Physics and Astronomy, Iowa State University, Ames, Iowa 50011, USA}

\author{P.~Yin}
\email[Corresponding author: ]{pengyin@iastate.edu}
\affiliation{College of Physics and Engineering, Henan University of Science and Technology, Luoyang 471023, China}
\affiliation{CAS Key Laboratory of High Precision Nuclear Spectroscopy, Institute of Modern Physics, Chinese Academy of Sciences, Lanzhou 730000, China}

\author{X.~Zhao}
\affiliation{Institute of Modern Physics, Chinese Academy of Sciences, Lanzhou 730000, China}
\affiliation{School of Nuclear Physics, University of Chinese Academy of Sciences, Beijing 100049, China}
\affiliation{CAS Key Laboratory of High Precision Nuclear Spectroscopy, Institute of Modern Physics, Chinese Academy of Sciences, Lanzhou 730000, China}

	\begin{abstract}
		Electric quadrupole (\textit{E}2) moments and transitions provide measures of nuclear deformation and related collective structure. However, matrix elements of the \textit{E}2 operator are sensitive to the nuclear wave function at large distances and are poorly convergent within the \textit{ab initio} no-core shell model approach. We perform for the first time a systematic study of the ratio of neutron to proton quadrupole transition matrix elements, $M_n/M_p$, with the no-core shell model. We find this observable is well-converged and provides a new and robust tool for comparing with experimental results. Our parameter-free results for $M_n/M_p$ for the carbon isotopes and $^{10}$Be compare well with experiment, where available, and provide additional support for 
		the emerging picture that \textit{ab initio} results 
		align well with collective enhancements in light nuclei.
	\end{abstract}
	
	\maketitle

\section{Introduction}
	Electromagnetic and strong interaction probes of elementary particles, atomic nuclei, atoms and molecules provide precision information on their structure and dynamics. Both static and transition properties over a range of allowed multipolarities can provide detailed insights into the underlying quantum structure and dynamics of these systems. Often, these probes can be treated as perturbative in character which facilitates comparison between theory and experiment.  On the other hand, for the applications in nuclear physics, it is very challenging to solve for the relevant quantities with sufficient predictive power for a quantitative comparison with experiment.  We show that, for \textit{ab initio} nuclear structure calculations, where convergence of individual long-range observables is problematic, ratios of selected observables, which involve  the neutron and proton quadrupole moments, are remarkably robust and well-suited both for comparing with experiment and for predicting yet-to-be measured properties. Such demonstrations provide guidance for developing suitable robust ratios in other fields of physics.
	
	Electric quadrupole (\textit{E}2) observables, such as electric quadrupole moments $Q_p$ where $p$ represents proton, and \textit{E}2 transition strengths $B(E2)$, reveal nuclear collective structure and 
dynamics~\cite{Bohr, DRowe}. Since \textit{E}2 observables are sensitive to the large-distance tails of the nuclear wave function, they are slowly convergent in \textit{ab initio} no-core shell model (NCSM) approaches~\cite{Bogner:2007rx, Maris:2013poa, Barrett2013, Odell:2016}. Even before convergence, useful predictions for static and dynamic \textit{E}2 observables have been derived by focusing on selected ratios~\cite{Calci_2016, Caprio:2019yxh, Caprio:2021,Caprio:2022pqp,Caprio:2022}, rather than the absolute scale of individual \textit{E}2 matrix elements. The prediction of $B(E2)$ values from these ratios is also found to be close to experiment data. On the other hand, in order to probe the nuclear quadrupole dynamics more completely, both proton and neutron quadrupole observables are needed and this is where strong interaction probes play a critical role~\cite{Bernstein:1983}.

	The dimensionless ratios  of \textit{E}2 observables considered in this work are defined in terms of matrix elements of the quadrupole operators:  
	\begin{gather*}
	\frac{M_n}{M_p} = \frac{\braket{J_f||\sum_{i \in n}r_{i}^2Y_2(\hat{r}_{i})||J_i}}{\braket{J_f||\sum_{i \in p}r_{i}^2Y_2(\hat{r}_{i})||J_i}},  \\
\frac{B(E2)}{Q_p^2} = \frac{5(J_s+1)\hat{J}_{s}^{2} (\hat{J}^{2}_{s}+2)
\braket{J_f||\sum_{i \in p}r_{i}^2Y_2(\hat{r}_{i})||J_i}^2}{16\pi J_s(\hat{J}_{s}^{2}-2)\hat{J}_{i}^{2}
\braket{J_s||\sum_{i \in p}r_{i}^2Y_2(\hat{r}_{i})||J_s}^2},  \\  
\frac{M_nQ_p}{M_pQ_n} = \frac{\braket{J_f||\sum_{i \in n}r_{i}^2Y_2(\hat{r}_{i})||J_i}\braket{J_s||\sum_{i \in p}r_{i}^2Y_2(\hat{r}_{i})|J_s}}{\braket{J_f||\sum_{i \in p}r_{i}^2Y_2(\hat{r}_{i})||J_i}\braket{J_s||\sum_{i \in n}r_{i}^2Y_2(\hat{r}_{i})||J_s}}.
\end{gather*} 
$J_i$, $J_f$ and $J_s$ represent the total angular momentum of, respectively, initial, final states and the state (initial or final) used to calculate the quadrupole moment~$Q_p$; 
$\hat{J}^{2}=2J+1$;
$r_i$ is the single-particle radius operator; 
 $n$~($p$) represents neutron (proton) contributions.
The sensitivity of \textit{E}2 observables to the large-distance properties of the nuclear wave function arises from the $r^2$ dependence of the quadrupole operators. This work serves as an extension to the class of ratios of $E2$ observables in Refs.~\cite{Caprio:2021,Caprio:2022,Caprio:2022pqp}. The ratio $B(E2)/Q_p^2$ was shown in Ref.~\cite{Caprio:2022pqp} to have a robust convergence, and it is reasonable to expect that $M_n/M_p$ and $M_{n}Q_p/M_pQ_n$ will also exhibit favorable convergence patterns in \textit{ab initio} calculations with limited basis spaces.

 In experiments, the $M_n/M_p$ ratio of the quadrupole transition from  2$_1^+$ to the 0$_1^+$ ground state  has been measured for several nuclei with different methods, such as $^{10}$Be (proton scattering~\cite{IWASAKI_2000}); 
$^{10}$C \big(proton scattering~\cite{Jouanne:2005pb}, $\rm^{10}C(\alpha,\alpha')$ reaction~\cite{Furuno_2019}\big);
$^{16}$C $\big(\rm{^{16}C}+{^{9}Be}$ scattering~\cite{Ong:2008},
${\rm{^{9}Be}({^{9}Be}},2p)$ reaction~\cite{Wiedeking:2008},
proton and deuteron scattering~\cite{Jiang_2020}\big);
$^{18}$O (hadron scattering~\cite{Bernstein_1979}); 
and $^{20}$O (proton scattering~\cite{JEWELL_1999}). 
The $M_n/M_p$ for the $5/2_1^+\to 1/2_1^+$ transition in $^{15}$C has
been determined via deuteron 
scattering~\cite{Chen_2022}. The $M_n/M_p$ ratios have received very limited theoretical attention: there are the $0\hbar\Omega$
shell model (SM)~\cite{Sagawa_2004,IWASAKI_2000,Yuan:2012NPA,Yuan:2012PRC-YSOX,Fortune:2016cop} 
and antisymmetrized molecular dynamics (AMD)~\cite{Kanada-Enyo:2011plo,Kanada-Enyo:2020wkr,Jouanne:2005pb} calculations
in carbon isotopes and $^{10}$Be; regarding the modern {\it ab initio} studies of~$M_n/M_p$,
there are only in-medium similarity renormalization group (IM-SRG) calculations for~$M_n/M_p$
in $^{16}$C~\cite{Jiang_2020} and our previous NCSM results for $^{15}$C~\cite{Chen_2022}  which we have
recalculated and present here with a higher precision. 

	\begin{figure*}
	\includegraphics[width=\textwidth]{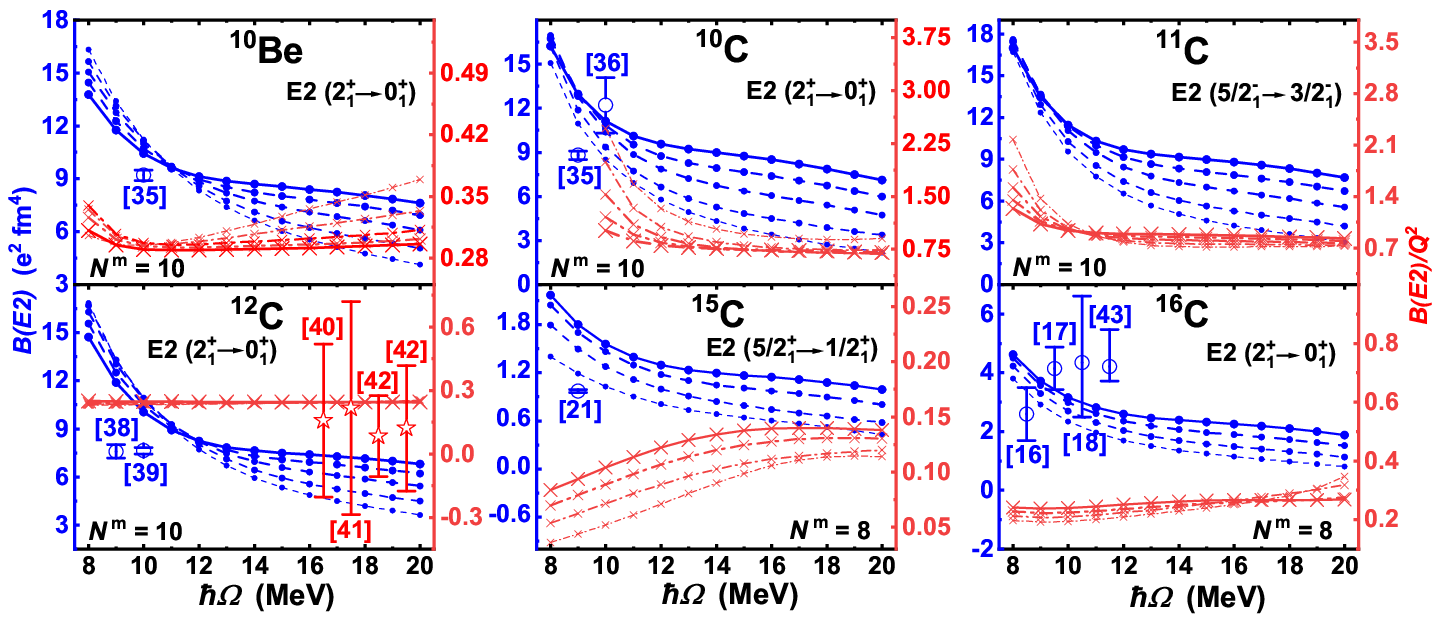}
	\caption{\label{fig:BE2} (Color online) $B(E2)$ (blue and scales to the left) and the ratios $B(E2)/Q_p^2$ (red and scales to the right) in $^{10}$Be, $^{10}$C, $^{11}$C, $^{12}$C, $^{15}$C, $^{16}$C for the indicated transitions are shown as functions of $\hbar\Omega$. NCSM results with Daejeon16 ${NN}$ interaction are shown by the dashed curves with  thickness increasing with~$N_{\max}$ for $N_{\max}= 2$, 4, 6 for $^{15,16}$C, and $N_{\max} = 2$, 4, 6, 8 for the other nuclei. The maximal $N_{\max}$ value is indicated at the bottom of each panel and is represented by the solid curves. For comparison, the available experimental $B(E2)$ and $B(E2)/Q_p^2$ 
	 data with their uncertainties are shown by open circles and stars respectively, with the  citation indicated in square brackets.
	}
	\end{figure*}

\section{Results and discussion}
In this work,  we demonstrate that the ratios of $M_n/M_p$ obtained from the \textit{ab initio} NCSM~\cite{Barrett2013, Navratil:2000gs,Navratil:2000ww} approach show good agreements with the available experimental data in the carbon isotopes up through $A=16$ and $^{10}$Be. Furthermore, the double ratio $M_{n}Q_p/M_pQ_n$, which represents the ratio of neutron and proton quadrupole matrix elements $M_n/M_p$ over the ratio of neutron and proton quadrupole moments $Q_{n}/Q_{p}$, shows even better convergence than the ratio of  $M_n/M_p$. We also find that the ratio of $B(E2)/Q_p^2$ in the carbon isotopes and $^{10}$Be displays excellent convergence. The combination of experimental $B(E2)$ values and converged theoretical $B(E2)/Q_p^2$ ratios can be used to predict proton quadrupole moments. We show that the resulting proton quadrupole moment predictions are remarkably close to the results of experiments. At the same time, it is possible to make a prediction of a neutron quadrupole moment by connecting the convergent theoretical results for $M_{n}Q_p/M_pQ_n$ with either theoretical  or experimental results for  $M_n/M_p$ and {$Q_{p}$}. 
	
			\begin{figure*}
		\includegraphics[width=\textwidth]{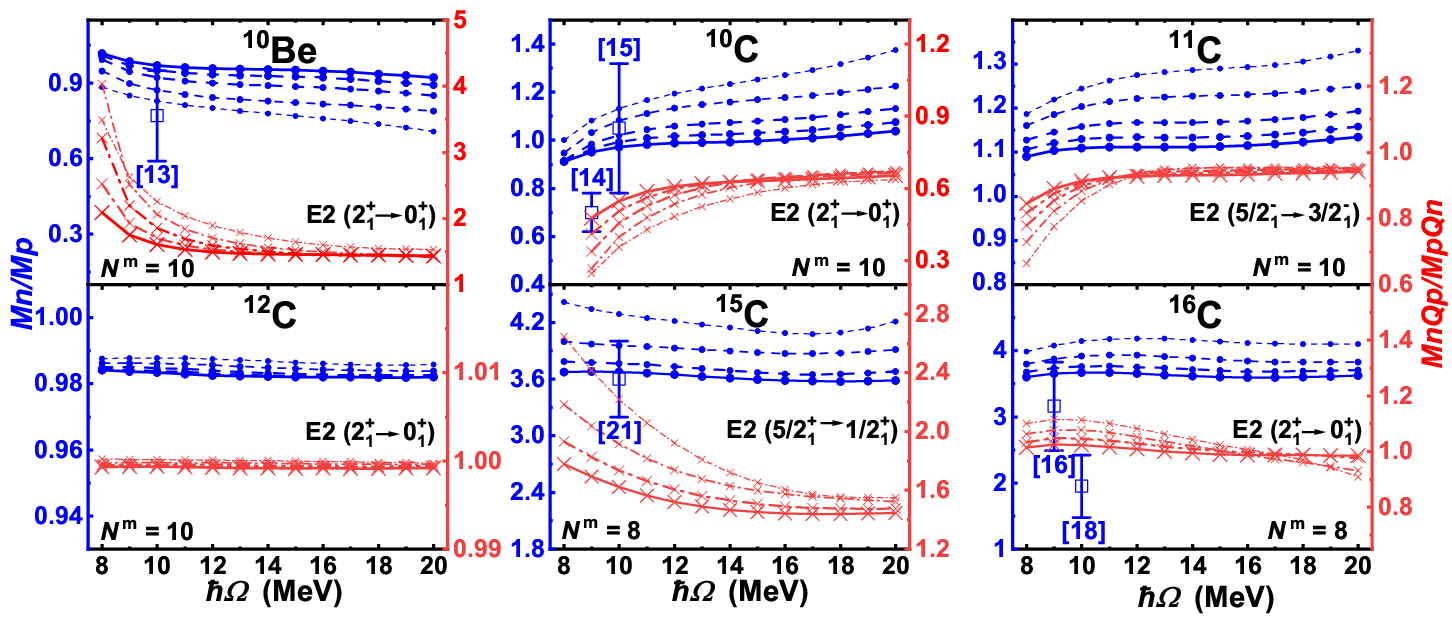}
		\caption{\label{fig:Mn-over-Mp} (Color online) Ratios $M_n/M_p$ (blue and scales to the left) and $M_{n}Q_p/M_pQ_n$ (red and scales to the right) in the same nuclei and for the same transitions as in  Fig.~\ref{fig:BE2}. See  Fig.~\ref{fig:BE2} for more details. Available experimental $M_n/M_p$ results 
 with their uncertainties are shown by open squares with the citation indicated in square brackets. 
}
	\end{figure*}

We perform the \textit{ab initio} NCSM calculations with the Daejeon16 $NN$ interaction~\cite{Shirokov:2016ead} for $^{10}$Be, $^{10}$C, $^{11}$C, and $^{12}$C up to $N_{\max} = 10$, $^{15}$C and $^{16}$C up to $N_{\max} = 8$, for which the experimental results of $M_n/M_p$ and/or proton quadrupole moment are available. This interaction is based on the Entem--Machleidt N$^3$LO chiral effective field theory  
interaction~\cite{Entem:2003ft}, softened via a similarity renormalization group 
transformation~\cite{Bogner:2006pc} so as to provide improved convergence, and then adjusted via a phase-shift equivalent transformation to better describe nuclei with $A\le16$. Using the MFDn code~\cite{MARIS:2010, Aktulga:2013}, we diagonalize the Hamiltonian of the system in a many-body harmonic oscillator basis  which is characterized by the basis energy scale $\hbar\Omega$ and the basis truncation parameter $N_{\max}$, the maximum number of oscillator excitation quanta allowed in the many-body space relative to the lowest Pauli-allowed configuration.

	In the NCSM approach, convergence is recognized when the calculated results become insensitive
	to increases in $N_{\max}$ and to variations in $\hbar\Omega$~\cite{Maris:2013poa}. 
	
	We first consider the ${E}2$ strength for the $2_1^+\to0_1^+$ transitions
in $^{10}$Be, $^{10}$C, $^{12}$C, $^{16}$C, $5/2_1^-\to 3/2_1^-$
in $^{11}$C, and $5/2_1^+\to1/2_1^+$ 
in $^{15}$C, shown in  Fig.~\ref{fig:BE2} with blue curves. While there is some tendency towards flattening of these curves with respect to $\hbar\Omega$ and compression of successive curves with respect to $N_{\max}$, the calculated values are still changing significantly with increasing $N_{\max}$. The dimensionless ratios of $B(E2)/Q_p^2$ 
yield the results shown in Fig.~\ref{fig:BE2} with red curves\footnote{In calculations of ratios~$B(E2)/Q_p^2$ and~$M_{n}Q_p/M_pQ_n$
	in $^{11}$C we use $Q_{p}$ and $Q_{n}$ of the $3/2^{-}$ ground
	state; in all other nuclei we use $Q_{p}$ and $Q_{n}$ of excited states since their ground states have
	no quadrupole moments.}. We find a near complete elimination of the $\hbar\Omega$ dependence at the highest $N_{\max}$ shown, as well as substantial compression of the curves for successive $N_{\max}$.

	The ratios of neutron and proton quadrupole matrix elements $M_n/M_p$ for our selected transitions are in Fig.~\ref{fig:Mn-over-Mp} with blue curves connecting calculated points. We note that the $\hbar\Omega$ dependence is less pronounced than for $B(E2)$ (Fig.~\ref{fig:BE2}). Moreover, the curves at higher $N_{\max}$ trend towards flattening while the spacing between curves for successive $N_{\max}$ decreases systematically.  These trends indicate a progression towards converged values which agree well with the available experimental results. We also present the $M_{n}Q_p/M_pQ_n$ ratios as red curves in Fig.~\ref{fig:Mn-over-Mp}.  Note that this ratio for $^{12}$C is almost
independent of $N_{\max}$ and $\hbar\Omega$ indicating a nearly complete convergence. For the transitions
in the remaining nuclei considered, the  
ratios of $M_{n}Q_p/M_pQ_n$ demonstrate a more significant $\hbar\Omega$ dependence at lower~$N_{\max}$,
but the curves rapidly compress and flatten at the higher $N_{\max}$ values shown.

    Based on the results displayed, 
     we make predictions for the quadrupole observables and summarize the results in Table.~\ref{tab:Res} in comparison with available experimental data
    in columns ex. For ratios of $B(E2)/Q_p^2$, $M_n/M_p$ and $M_{n}Q_p/M_pQ_n$, we provide converged results (columns con) which are obtained in the following manner.
    First, we consider 5 MeV intervals of $\hbar\Omega$ values, hereafter referred to as windows, for the calculated ratios where the curves of the largest $N_{\max}$ results in each panel are approximately flat, i.\,e., the  $\hbar\Omega$ dependence is the weakest.
    For the ratios of $B(E2)/Q_p^2$, the windows are set from 11 
to 16~MeV for $^{10}$Be,  and from 
15 to 20~MeV for the carbon isotopes. 
For the ratios of $M_n/M_p$, the windows are set from 11 to 16~MeV for $^{10}$Be, $^{10}$C, $^{11}$C,  and from 15 to 20~MeV for $^{12}$C, $^{15}$C, $^{16}$C. The windows are fixed from 15  to 20~MeV for the ratio of $M_{n}Q_p/M_pQ_n$ for all cases. Then we calculate the differences between results at the largest and the second largest $N_{\max}$ at different $\hbar\Omega$ values in this window. Finally, the computed result at the largest $N_{\max}$ corresponding to the minimum of this difference is selected as our final result, while the maximum of this difference is treated as the first part of the uncertainty $\Delta$1. The second part of our uncertainty, 
$\Delta2$, is from the magnitude of the difference between the minimum and maximum of results from the largest $N_{\max}$ in the selected $\hbar\Omega$ window. 
 We adopt the sum of uncertainties of converged ratios, $\Delta1 + \Delta2$, as the total uncertainty. Taking experimental and theoretical uncertainties into consideration, we show that the theoretical predictions (denoted by con) of $M_n/M_p$ are in reasonable agreement with experimental data in Table~\ref{tab:Res}.
    
		\begin{table*}[htbp]
		
		\begin{adjustbox}{angle=90}
			
			\begin{minipage}[\textheight]{1.4\textwidth}

				\caption{\label{tab:Res} 
					Predicted and experimental $B(E2)$, $B(E2)/Q_p^2$, $M_n/M_p$,  $M_nQ_p/M_pQ_n$, $Q_p$, and $Q_n$ are listed along with their uncertainties. The columns labeled as ex, cr,
					con and th+ex present respectively the experimental data, crossing point values, converged values and predictions from the combination of experimental and theoretical results. The column other-th presents the results from previous calculations of $M_n/M_p$ 
					(\textit{ab initio} IM-SRG~\cite{Jiang_2020}, 
					AMD~\cite{Jouanne:2005pb,Kanada-Enyo:2011plo,Kanada-Enyo:2020wkr},
					$0\hbar\Omega$ SM calculations with 
					Millener--Kurath~\cite{IWASAKI_2000,Sagawa_2004,Yuan:2012NPA}, 
					Warburton--Brown~\cite{Yuan:2012NPA}, 
					Yuan--Suzuki--Otsuka--Xu~\cite{Yuan:2012PRC-YSOX} interactions and using the SM version 
					suggested in Ref.~\cite{Fortune:2016cop}.)}
				
				\begin{tabular}{c|cr|c|lll|l|llll|lll}
					\toprule
					\toprule
					\multicolumn{1}{c}{\space} & \multicolumn{2}{c}{$B$($E$2) ($e^2$~fm$^4$)} & \multicolumn{1}{c}{$B$($E$2)/$Q_{p}^2$}&\multicolumn{3}{c}{$M_n/M_p$} & \multicolumn{1}{c}{$M_nQ_p/M_pQ_n$} & \multicolumn{4}{c}{$Q_{p}$ ($e$~fm$^2$)} & \multicolumn{3}{c}{$Q_{n}$ (fm$^2$)}   \\
					\cmidrule(rl){2-3} \cmidrule(rl){4-4} \cmidrule(rl){5-7}\cmidrule(rl){8-8} \cmidrule(rl){9-12} \cmidrule(rl){13-15}
					\multicolumn{1}{c}{\space} & \multicolumn{1}{c}{ex} & \multicolumn{1}{c}{cr}& \multicolumn{1}{c}{con}  & \multicolumn{1}{c}{ex}& \multicolumn{1}{c}{con} & \multicolumn{1}{c}{other-th} & \multicolumn{1}{c}{con} & \multicolumn{1}{c}{ex} & \multicolumn{1}{c}{$\displaystyle\frac{B^{ex}(E2)}{B(E2)/Q_p^2}$} & \multicolumn{1}{c}{$\displaystyle\frac{B^{cr}(E2)}{B(E2)/Q_p^2}$} & \multicolumn{1}{c}{cr} &\multicolumn{1}{c}{th+ex} & \multicolumn{1}{c}{$\displaystyle\frac{Q_p^{cr}({M_n}/{M_p})^{\rm con}}{M_nQ_p/M_pQ_n }$} & \multicolumn{1}{c}{cr}  \\
					
					
					\midrule
					\midrule
					$^{10}$Be & 9.2(3)~\cite{McCutchan:2012}
					&9.48(2) & 0.29(1) & 1.1(1)~\cite{IWASAKI_2000}
					& 0.96(4) & 0.62~\cite{IWASAKI_2000} & \hspace{2.4ex}1.45(27)& &$\hspace{1ex}-5.63(5)$
					&$\hspace{1ex}-5.72(3)$ &$-5.81(2)$
					& $-4.3(2)$
					&$\hspace{4ex}-3.8(2)$&
					
					\\
					(2$_1^+$\textrightarrow  0$_1^+$)& 
					& & & 
					& &0.88~\cite{Kanada-Enyo:2020wkr} &  & & 
					& &
					& 
					&&
					
					\\[2mm]
					$^{10}$C &8.8(3)~\cite{McCutchan:2012}
					&12.7(8)\hspace{1.2ex} & 0.76(6)
					&0.70(8)~\cite{Jouanne:2005pb}
					&0.98(5) & 1.29~\cite{Sagawa_2004}&\hspace{2.4ex}0.62(15) 
					&	&	$\hspace{1ex}-3.41(9)$
					& $\hspace{1ex}-4.1(1)$	& &$ -5.8(4)$ &  &$\hspace{1ex}-5.9(2)$
					
					\\
					(2$_1^+$\textrightarrow  0$_1^+$) & 12.2(19)~\cite{Fisher-1968zza}
					& & & 1.05(20)~\cite{Furuno_2019}
					& &1.11~\cite{Jouanne:2005pb} & 
					&	 &	$\hspace{1ex}-3.9(2)$
					&	&
					&$-4.4(3)$
					&  & 
					
					\\
					\space & 
					& &  & 
					& & 1.3~\cite{Kanada-Enyo:2011plo} & 
					&	 &	
					&	&
					&
					&  &
					
					\\[2mm]
					$^{11}$C& & 12(1)\hspace{3ex} & 0.94(8)
					& &	1.09(5) & 1.19~\cite{Jouanne:2005pb}& \hspace{2.4ex}0.93(4) &3.33(2)~\cite{Stone:2016bmk} &	&\hspace{3ex}3.6(2) & &\hspace{2ex}3.9(1) &  & \hspace{3ex}3.93(2)
					\\
					($\frac{5}{2}_1^-$\textrightarrow  $\frac{3}{2}_1^-$) & 
					& & &
					& &  & 
					&	 &	
					&	&
					&
					&  &
					
					\\[2mm]
					$^{12}$C & \hspace{2ex}7.63(19)~\cite{D'Alessio_2020}
					& 8.08(4) & 0.24(1)
					&	& 0.982(2) &1~\cite{Sagawa_2004} & \hspace{2.4ex}0.9992(3)	&5.97(30)~\cite{D'Alessio_2020}	& \hspace{3ex}5.59(4)
					&\hspace{3ex}5.76(5) & \hspace{2ex}5.7(2)
					& \hspace{2ex}5.9(7)& \hspace{6ex}5.6(2) & \hspace{3ex}5.6(2)
					
					\\
					(2$_1^+$\textrightarrow  0$_1^+$) &\hspace{2ex}7.94(66)~\cite{Pritychenko:2016}
					& & &
					& & & 
					&\hspace{.6pt}6(3)~\cite{VERMEER:1983}	 &	\hspace{3ex}5.71(9)
					&	&
					&\hspace{2ex}5.9(5)
					&  & 
					
					\\
					\space & 
					& & &
					& & & 
					&5.3(44)~\cite{KUMARRAJU:2018}	 &
					&	&
					&\hspace{2ex}5.2(8)
					&  & 
					
					\\
					\space & 
					& & &
					& & & 
					&9.5(18)~\cite{SAIZLOMAS:2023}	 &
					&	&
					&\hspace{2ex}9.3(2)
					&  &

					\\[2mm]
					$^{15}$C & \hspace{.8ex}0.97(2)~\cite{Chen_2022}
					& & 0.14(2) & 3.6(4)~\cite{Chen_2022} &3.6(2) & & \hspace{2.4ex}1.44(29) & & $\hspace{1ex}-2.7(3) $&& $-3.11(6)$
					& $-6.7(3)$ & $\hspace{4ex}-7.8(2)$ & $\hspace{1ex}-8.3(5)$
					\\
					($\frac{5}{2}_1^+$\textrightarrow  $\frac{1}{2}_1^+$) & 
					& & &
					& & & 
					&	 &
					&	&
					&
					&  &

					\\[2mm]
					$^{16}$C & 2.6(9)~\cite{Ong:2008}
					& & 0.27(4) & 3.17(67)~\cite{Ong:2008}
					&3.6(2) & 2.54~\cite{Jiang_2020}& \hspace{2.4ex}0.99(6)
					&	 &	$\hspace{1ex}-3.12(9)$
					&	&$-3.5(8)$
					&$-10(2)$
					& $\hspace{3ex}-12.7(2)$ & $-12.4(5)$

					\\
					(2$_1^+$\textrightarrow  0$_1^+$) &  \hspace{2ex}4.15(73)~\cite{Wiedeking:2008}
					& &  & 2.4~\cite{Wiedeking:2008}
					& &  6.07~\cite{Sagawa_2004}&
					&	 &	$\hspace{1ex}-3.9(2)$
					&	&
					&$-9.5(3)$
					&  & 
					
					\\
					\space &  \hspace{2.5ex}4.34\mbox{$_{-1.85}^{+2.27}$}~\cite{Jiang_2020}
					& &  & 1.95(47)~\cite{Jiang_2020}
					& & 2.11~\cite{Yuan:2012NPA} &
					&	 &	$\hspace{1ex}-4.0(5)$
					&	&
					&$-7.9(3)$
					&  & 
					
					\\
					\space &  \hspace{2.3ex}4.21\mbox{$_{-0.50}^{+1.26}$}~\cite{Petri_2012}
					& & &
					& & 2.47~\cite{Yuan:2012NPA} &
					&	 &$\hspace{1ex}-3.9(3)$
					&	&
					&
					&  &
					
					\\
					\space & 
					& & &
					& &  2.48~\cite{Yuan:2012PRC-YSOX}&
					&	 &
					&	&
					&
					&  & 
					\\
					\space & 
					& & &
					& & 3.0~\cite{Fortune:2016cop} &
					&	 &
					&	&
					&
					&  &
					
					\\
					\space & 
					& & &
					& & 4.19~\cite{Kanada-Enyo:2011plo} &
					&	 &
					&	&
					&
					&  &
					\\
					\bottomrule
					\bottomrule
				\end{tabular}
				
			\end{minipage}
			
		\end{adjustbox}
		
	\end{table*}
	
	 In the NCSM calculations, a crossing point is sometimes found for observables while varying $N_{\max}$ and $\hbar\Omega$. Following the practice of Refs.~\cite{Caprio:2021,Caprio:2022,Shirokov:2016ead}, we cite the crossing point of the curves obtained with $N_{\max} = N^m$ and $N_{\max} = N^{m}-2$, where $N^m$ is the largest attainable\- $N_{\max}$ value for the given nucleus,  as our 
estimates for $B(E2)$, $Q_p$ and $Q_n$ and mark them as cr in Table I. The difference between the observable values at this crossing point and that  obtained with $N_{\max}=N^{m}-2$ and $N_{\max} = N^{m}-4$ is cited as our rough estimation of the uncertainty.
The resulting ${E}2$ strength estimations of $2_1^+\to 0_1^+$ transition in $^{10}$Be and $^{12}$C are close to the experiments, while in $^{10}$C the predicted $B(E2)$ value is consistent with an old experiment~\cite{Fisher-1968zza} but is approximately 30\% higher than suggested by a more recent
experiment~\cite{McCutchan:2012}. The predictions for $B(E2{:}\,2_1^+ {\to}\, 0_1^+$) from other \textit{ab initio} approaches (in ${e}^2$~fm$^4$), such as 8.8(4) from Green’s Function Monte Carlo (GFMC)~\cite{McCutchan:2012} and 9.29 from No-core Monte Carlo Shell Model 
\mbox{(NCMCSM)}~\cite{Liu:2018cer} for $^{10}$Be, 15.3(1.4) (GFMC)~\cite{McCutchan:2012} and 9.30 \mbox{(NCMCSM)}~\cite{Liu:2018cer} for $^{10}$C, 7.65 (NCMCSM)~\cite{Otsuka:2022bcf} and 8.68(79) In-medium NCSM (IMNCSM)~\cite{D'Alessio_2020} for $^{12}$C are in reasonable agreement with our results. 

	 We have not observed the crossing points of $B(E2)$ for $^{15}$C and $^{16}$C with current results, which may suggest  that these crossings might appear at $\hbar\Omega$ and/or $N_{\max}$ values beyond the ranges of our investigation. The combination between prediction of $B(E2)/Q_p^2$ (marked as con) and experimental or theoretical $B(E2)$ values gives 
an alternative  prediction of Q$_p$ marked respectively by $\displaystyle\frac{B^{ex}(E2)}{B(E2)/Q_p^2}$ or $\displaystyle\frac{B^{cr}(E2)}{B(E2)/Q_p^2}$
in Table~\ref{tab:Res}. 

The junction of $Q_p$, $M_n/M_p$ and $M_{n}Q_p/M_pQ_n$ estimations was adopted to predict another quadrupole observable, $Q_n$, which is essential 
for the understanding of proton-neutron asymmetry in nuclear deformation, but is challenging to measure in experiment. 
The results in the $Q_n$ column labeled as th+ex were predicted from the combination of experimental (columns ex) and theoretical data. Specifically, we used the experimental
value for $M_n/M_p$, proton quadrupole moment $Q_p$ (column $\displaystyle\frac{B^{ex}(E2)}{B(E2)/Q_p^2}$)  
and our  $M_{n}Q_p/M_pQ_n$ results to make the predictions
for 
the neutron quadrupole moments for $^{10}$Be, $^{10}$C, $^{15}$C and $^{16}$C, while  
our predictions of  $Q_{n}$ for $^{11}$C and $^{12}$C in the column th+ex were obtained by making use of 
$M_n/M_p$ (column~con), experimental $Q_{p}$ and our $M_{n}Q_p/M_pQ_n$ values. 
The $Q_{n}$ predictions in the column $\displaystyle\frac{Q_p^{cr}({M_n}/{M_p})^{con}}{M_nQ_p/M_pQ_n }$ were obtained utilizing the theoretical $M_n/M_p$  (column con) and $Q_{p}$ 
(column cr) results. 

The theoretical predictions for $Q_p$ and $Q_n$ by converged ratios are close to those obtained from
the crossing points (columns~cr). 
The estimations of $Q_p$ by different methods agree well with the presented experimental results for the 3/2$_1^-$ level in $^{11}$C and 2$_1^+$ level in $^{12}$C, which provides a support for our predictions of $Q_p$ and $Q_n$.
	
\section{conclusion}
	In conclusion, ratios of $M_n/M_p$ obtained from the \textit{ab initio} NCSM approach show good agreement  with available experimental data. We also provide robust \textit{ab initio} calculations for ratios of $B(E2)/Q_p^2$ and $M_{n}Q_p/M_pQ_n$. The predicted proton quadrupole moments from the converged ratio results are close to the available experimental results. Although the neutron quadrupole moment itself is not directly accessible experimentally, \textit{ab initio} calculations can now provide robust predictions that lead to insights in the  neutron-proton asymmetry in quadrupole deformation. By demonstrating that ratios of weakly convergent theoretical results provide robust means for comparing theory with experiment, we hope to stimulate similar investigations in other areas of physics.
	
	Finally, we note that in some nuclei discussed here, the experimental $B(E2)$ results
exhibit collective behavior since they are significantly enhanced from unity  when expressed in Weisskopf units (W.u.): $^{10}$Be (7.19(23) W.u.~\cite{McCutchan:2012}), $^{10}$C (6.87(23) W.u.~\cite{McCutchan:2012}, 9.5(15) W.u.~\cite{Fisher-1968zza}) and $^{12}$C (4.67(12) W.u.~\cite{D'Alessio_2020}, 4.86(41) W.u.~\cite{Pritychenko:2016}). The \textit{ab initio} theoretical $B(E2)$ results from GFMC~\cite{McCutchan:2012}, IMNCSM~\cite{D'Alessio_2020}, NCMCSM~\cite{Liu:2018cer,Otsuka:2022bcf}, and our NCSM also support this enhancement. The agreement between theory and experiment for these enhancements as well as for the $M_n/M_p$ results in these same nuclei support the assertion that {\it ab initio} nuclear structure can describe accurately the quadrupole dynamics of light nuclei. The possibility to describe collectivity by {\it ab initio} approaches is also supported by the description of rotational bands within NCSM~\cite{Caprio:2019yxh, Caprio:2021}.

\begin{acknowledgements} 
	We thank J. Chen and J. Li for useful discussions. J.~P.~Vary is supported by the US Department of Energy under Grant No.~DE-SC0023692. A.~M.~Shirokov is supported by Chinese Academy of Sciences President’s International Fellowship Initiative Grant No.~2023VMA0013. H.~J.~Ong is supported by the Chinese Academy of Sciences “Light of West China” Program, the National Natural Science Foundation of China under Contract Nos.~12175280 and 12250610193. X.~Zhao is supported by new faculty startup funding by the Institute of Modern Physics, Chinese Academy of Sciences, by Key Research Program of Frontier Sciences, Chinese Academy of Sciences, Grant No.~ZDB-SLY-7020, by the Natural Science Foundation of Gansu Province, China, Grant No.~20JR10RA067, by the Foundation for Key Talents of Gansu Province, by the Central Funds Guiding the Local Science and Technology Development of Gansu Province, Grant No.~22ZY1QA006, by international partnership program of the Chinese Academy of Sciences, Grant No.~016GJHZ2022103FN, by National Natural Science Foundation of China, Grant No. 12375143, by National Key R$\&$D Program of China, Grant No.~2023YFA1606903 and by the Strategic Priority Research Program of the Chinese Academy of Sciences, Grant No.~XDB34000000. P.~Yin is supported by the Gansu Natural Science Foundation under Grant No.~23JRRA675. This research is supported by Gansu International Collaboration and Talents Recruitment Base of Particle Physics (2023–2027), and supported by the International Partnership Program of Chinese Academy of Sciences, Grant No.~016GJHZ2022103FN. A portion of the computational resources were also provided by Gansu Computing Center and Gansu Advanced Computing Center.
\end{acknowledgements}

\bibliography{Refs_E2ratios}
\end{document}